# Self-Oscillatory Light Emission in Plasmonic Molecular Tunnel Junctions

Riccardo Zinelli[a], Zijia Wu[a], Christian A. Nijhuis[a*], Qianqi Lin[a*]

[a] Hybrid Materials for Opto-Electronics Group, Department of Molecules and Materials, MESA+ Institute for Nanotechnology and Center for Brain-Inspired Nano Systems, Faculty of Science and Technology, University of Twente, P.O. Box 2017, 7500 AE Enschede, The Netherlands

*Author to whom correspondence should be addressed:
c.a.nijhuis@utwente.nl and q.lin@utwente.nl

**Abstract**

Self-oscillators are intriguing due to their ability to sustain periodic motion without periodic stimulus. They remain rare as achieving such behavior requires a balance of energy input, dissipation and non-linear feedback mechanism. Here, we report a molecular-scale optoelectronic self-oscillatory system based on electrically excited plasmons. This system generates light *via* inelastic electron tunnelling, where electrons lose their energy to molecules and excite the surface plasmon polaritons that decay radiatively. Time-series imaging of photon emission in gold–naphthalene-2-thiol–EGaIn junctions, together with correlation mapping of individual emission spots, reveal long-lived (~1000 s), low-frequency oscillations (1–20 mHz) interspersed with transient high-frequency (20-200 mHz) bursts. This behavior can be explained by attributing individual emission spots to single-molecule resistors that follow Kirchhoff's circuit laws. Induced by tunnelling current, these individual spots emit in a correlated way, self-sustaining the overall oscillatory emission from the whole junction. Our observation is of great interest as it resonates with a broader understanding of similar molecular-scale dynamic systems such as picocavities, offering exciting potential for optoelectronic and sensing applications.

**Introduction**

Oscillations are ubiquitous in nature and important for a wide range of biological,[1–3] physical,[4] and chemical systems,[5,6] to model blood flow with microfluidic networks,[7] soft actuators for soft robotics applications[8,9], or to develop oscillatory networks for brain-inspired computing[10,11]. It is challenging to develop and understand self-oscillators (oscillations without periodic external stimuli)[12,13]. These self-oscillators exhibit rich and time-dependent dynamics that often found in natural systems (such as bursting, edge-of-chaos or chaotic behavior). To sustain self-oscillations, at least two physical processes are needed that are coupled (with a feedback mechanism), which must continuously balance energy input and dissipation.[14] This is particularly challenging to achieve in nanoscale opto-electronic devices, for example, where Joule heating and very large optical and electrical fields (GV/m) are routinely encountered.[15] Here we report molecular-scale optoelectronic self-oscillators based on plasmonic molecular tunneling junctions with surprisingly complex oscillatory behavior. These oscillators require only a constant voltage input that results in current flow and oscillatory plasmon-excitation followed by photon emission. The oscillatory nature of this system stems from the coupling of current flow to structural changes (adatom migration and molecular conformations[16,17] similar to those proposed in other plasmonic structures subjected to high fields, *e.g.,* "picocavities" in surface- and tip- enhanced Raman spectroscopy)[18–20] by Kirchhoff circuit rules.[21] This oscillatory interplay of nano-electronics and plasmonics potentially offers a complementary approach to widely studied static nano-optoelectronic devices, but our findings also give new insights into the dynamic behavior of molecular-scale junctions.

The interplay between the high electromagnetic fields and the dynamical behavior in plasmonic nanocavities remains underexplored. On one hand, optically driven nanocavities with metal-molecule interfaces constructed from nanoparticle-on-mirror,[22] and nanoparticle dimers,[23] have been widely studied and found applications especially in SERS.[24] Such systems show rich dynamics involving adatom formation coupled with optically-excited plasmons leading to signal enhancement interesting for single-molecule spectroscopy (*i.e.*, picocavities).[20,25] On the other hand, electrically driven tunnelling junctions with two planar electrodes,[16] or with a scanning probe as top electrode,[26] excite plasmons *via* inelastic electron tunnelling (IET).[27] Here, electrons tunnel through molecules and excite the surface plasmon polaritons (SPP) of which some radiatively decay resulting in photons that can be detected in the far field. Such electrically driven junctions may also show dynamical behavior related to molecular conformations,[16,17,28] but, in contrast to optically driven junctions, protrusions induced by adatoms may have deleterious effects in molecular based electronic devices because of the risk of filament formation leading to shorts.[29]

Our molecular-scale self-oscillatory system is based on electrically-excited plasmons in the tunnelling junctions of self-assembled monolayers (SAMs) derived from 2-naphthalenethiols (NPT, see Fig. 1 below). Similar junctions have been reported before, where the bottom electrode of Au supports the SAM, which is then contacted with EGaIn (eutectic gallium indium alloy) top-electrodes.[16,30–32] In these junctions, the light emission originates from discrete diffraction-limited spots with photon emission statistics similar to point sources can be controlled *via* the applied bias voltage.[16] The molecular properties defines the shape of the tunneling barrier where functional or polarizable groups affect the tunneling barrier height and electrostatic potential profile of the junctions,[33,34] and the molecular length the tunneling barrier width[16]. The light emission originates from discrete spots because of the surface roughness of the electrodes[35,36] and the sensitivity of the tunneling probability of charge carriers with distance.[33,37] Especially the EGaIn top-electrode is rough leading to a very small effective contact of about $10^6$ times smaller than the geometrical footprint of the EGaIn tip with the SAM making it possible to study single-molecule events.[16,28,32] For example, by changing the molecular tilt angle directional SPP excitation has been demonstrated[33] and intermittent light emission could be controlled by changing the rigidity of the molecular back bone.[17]

Here we show that the emission spots in plasmonic EGaIn molecular junctions are correlated leading to oscillatory emission behavior. We propose a feedback mechanism driven by dynamic changes in the local current pathways which we attribute to molecular conformational changes and adatom formation in the junction. These structural fluctuations modulate current flow and inelastic tunneling events underlying the feedback loops responsible for the observed self-oscillations. Since the overall potential drop across the junction remains constant (following Kirchhoff circuit rules), all emission spots are coupled where the change in voltage drop across one spot affects the drop across all other spots. This configuration effectively couples electrical inputs to optical oscillating outputs, paving the way for a novel class of nano-scale optoelectronic oscillators.

**Results and discussion**
**Molecular tunnelling junctions.**

Figure 1a shows a schematic illustration of the junctions with NPT SAMs fabricated using the cone-shaped tip EGaIn technique[38] or with EGaIn stabilized in microfluidic devices.[16,39] Briefly, we formed the SAMs on clean, flat template-stripped Au electrodes and formed electrical contact with non-invasive cone-shaped tips of EGaIn.[38] The surface quality of both freshly template stripped Au electrodes and the NPT samples were imaged using atomic force microscopy (AFM), giving a root-mean-squared surface roughness of 0.33 nm and 0.42 nm, respectively, both measured over an area of $1 \times 1$ μm$^2$ (Fig. S1).

The measured current density $J$ (A/cm$^2$) is described by the general tunneling equation,

$$J = J_0\, e^{-\beta d} \quad (1)$$

where $d$ is the barrier width (in our case the thickness of the NPT monolayer), $\beta$ the tunneling decay coefficient (that depends on the barrier height) and $J_0$ is the pre-exponential factor. To confirm the stability of the junctions, we performed $J(V)$ characterization by measuring 26 junctions in a bias window of -0.5 to 0.5 V in steps of 50 mV, for a total of 500 scans. Measurements at positive bias were limited due to the wind force effect, which induces migrations of atoms of the electrode material and the subsequent formation of filaments, causing electrical shorts.[40] We plotted the log-values of the currents in histograms to which fitted Gaussians to obtain the Gaussian mean of $\log_{10}|J|$ vs. $V$ curve shown in Fig, 1c (see Fig. S3 for more details). The shape of $J(V)$ is expected for coherent tunnelling and also the currents are similar to previously reported values for similar molecules, such as thiols with biphenyl backbones.[33,41]

To visualize the interface of the molecular junctions and to confirm we have molecular plasmonic tunneling junctions, we recorded the optical images using an electron multiplying charge-coupled device (EMCCD) camera at an applied bias of -2.2 V (Fig. 1b, also see Video S1). In agreement with earlier work, the contact area is inhomogeneous where the area with the EGaIn top-electrode appears as lighter grey scale than area of poor contact.[14] The emission events are all in areas with good top-electrode contacts clearly visible as bright spots. To ensure the molecular junctions excite plasmons, we recorded the emission spectra as function of $V$, using EGaIn constrained in a microfluidic network in polydimethylsiloxane (PDMS) rubber which can readily placed on our inverted microscope platform, using well established procedures[15,17,39]. Fig.1d shows these spectra consist of two peaks associated with a localized surface plasmon (LSP) mode and a propagating surface plasmon polariton (SPP) as published before.[16] The peak associated with the SPP mode blue-shifts as function of $V$ following the quantum cut-off law (Fig. S2) as expected.[16,17,33,34] From these experiments we conclude that we have high quality plasmonic molecular tunneling junctions where plasmon are excited *via* inelastic tunneling through individual molecules (as depicted in Fig. 1a) as previously reported.

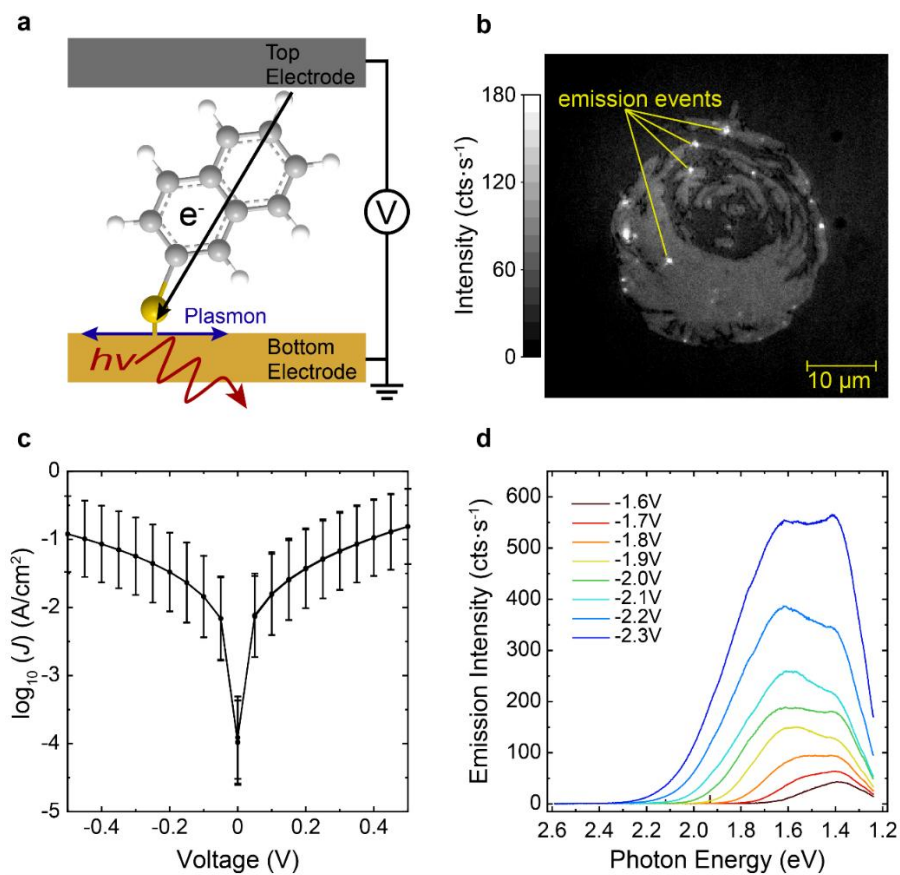

**Fig. 1 Au-NPT//EGaIn junctions electrical and spectral characterization of molecular tunneling junction. a** Schematic illustration of the Au-NPT//EGaIn junctions, where "-" indicates the metal-molecule, and "//" the noncovalent interface of the SAMs with the EGaIn electrode. **b**

EMCCD image (viewed through bottom electrode) of a Au-NPT//EGaIn junction at -2.2 V. Optical intensity is in grey scale. **c** Log-average $J(V)$ curves for junctions, measured in ± 0.5 V range with a 50 mV step size. **d** Plasmonic emission spectra of junctions obtained at the indicated voltages.

**Dynamics of electrically excited plasmons.**

According to previous works[16,17,33], the emission intensity from the junction linearly scales with the current across the junction (. Therefore, the intensity of individual emission spots should also depend on the current that flows across an emission sport. In principle, the current can vary with the tunneling distance or when the molecule-electrode coupling changes. The molecule-electrode coupling can change when the molecule undergoes conformational changes which can be induced by inelastic tunneling events coupled to the vibrational modes of the molecule[15–17,42–44], or by changes in the applied electric field[15,16]. In case the molecule makes a good contact with the electrode or stands up to better align with the electric field, electron tunnelling is more favorable increasing the current, IET, and emission, compared to the case when the molecule points away from the electrode (Fig. 2a).[33] This coupling between current and conformation of the molecule results in transient emission of each spot. Besides molecular considerations, the effective tunneling distance can also be reduced when adatoms (Fig. 2b) or small clusters of adatoms form. Here, the effective change in $d$ leads to an exponential increase in $J$ (Eq. 1) and thus also in IET events and associated light emission. It is known that adatom formation (Fig. 2b) enhances the local electromagnetic field by 3-10 times additional to the enhancement factor in nanocavities of 100-300.[19,45,46] Adatoms and clusters of gold adatoms are ubiquitous in thiolate SAMs due to vacancy islands arising from lattice strain[47]. There is no straightforward way to discriminate between these two proposed mechanisms in molecular tunnelling junctions to date, but conformational changes will most likely lead to regular transient (*i.e.,* short-lived) events with high light intensities while adatoms or cluster formation may result in picocavities or "flares"[25,48,49] characterized by very bright flashes.

These effects are readily visible from Video S2 (also Fig. S4) showing a junction that initially was stable but then shorted after 183 s. "Stable" here and below refers to a stable current and emission intensity from the overall junction, regardless of dynamic fluctuations in emission from individual spots. The video shows that emission spots appear when the applied voltage is ramped to -2.2 V and emission remains stable. At $t$ = 60 s in video S1, at -1.5 V bright spots start to appear and increase in emission intensity until the voltage reaches -2.2 V (Fig. S4a), until the junctions shorts at the end as the extreme case, with bright emission from the shorting area of the junctions (Fig. S4d). This visualization supports our interpretation of bright spots as flares caused by movement of adatom clusters, but also highlights the dynamics of the emission spots of a stable junction despite that the current (and associated voltage drop) across the junction remains constant, until the junction starts to break down (Fig. S4c).

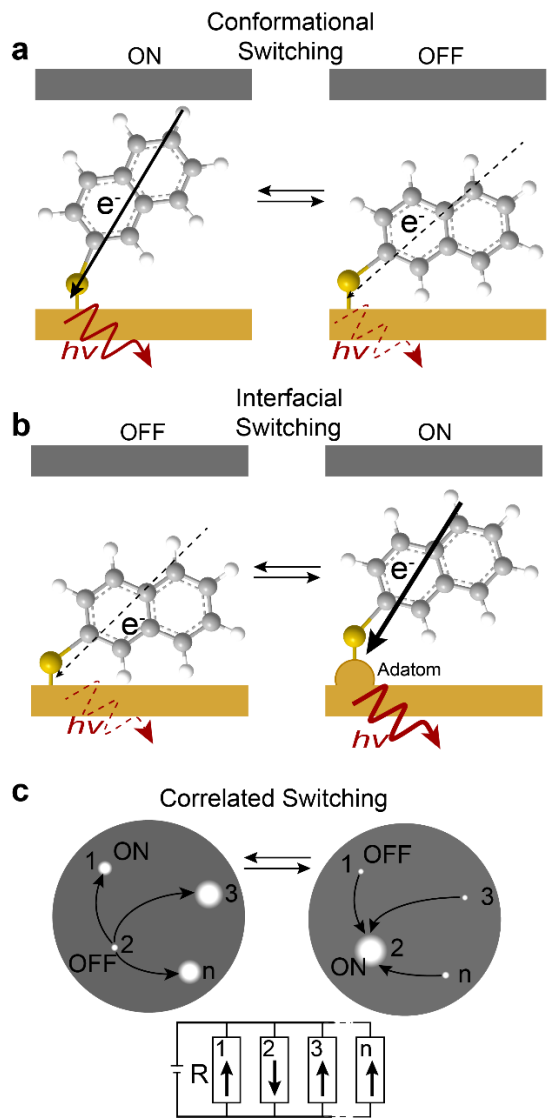

**Fig. 2 Dynamic junction emission based on molecular configuration switching.** Schematic representations of **a** molecular conformational and **b** interfacial changes due to adatom formation, as origins of the dynamical behavior of the junctions. Solid black arrow: Electron tunneling through the junction. Dashed black arrow: Electron tunneling decreases as the molecule lies down so becomes less aligned with the electric field. Solid red arrow: Strong light emission. Dashed red arrow: Weak or absent light emission. **c** Schematic illustration showing a dynamic plasmonic molecular junction along with a Kirchhoff's circuit, where each emission spot is a time-varying resistor.

Since the emission events are dynamic and occur at discrete and localized spots, these spots are, in principle, coupled *via* the electric field since the total current that flows across a stable junction remains constant. Following this line of thought, these emission spots can be regarded as time-varying resistors connected in parallel, forming a Kirchhoff circuit (Fig. 2c). In other words, when the current drops in one or more spot(s), other spot(s) has to increase to maintain a constant overall current, providing a coupling mechanism between the emission spots. Since current and photon emission rates scale linearly in stable junctions, recording the dynamics of emission spots can be directly related to current flow (and associated voltage drop), and to establish whether the emission spots in the junction behave as Kirchhoff circuit elements. The next sections show that this is indeed the case leading to complex oscillatory behavior.

**Oscillations of a single point source.**

To study the dynamic behavior of individual emission events from NPT molecular junctions, we recorded a series of EMCCD images with integration time of 500 ms at a constant bias of $V = -2.2$ V that was dominated by one spot (Fig. 3 and Video S3). We followed the emission events for 900 s, after which we stopped the experiment. Fig. 3a shows the 3D image stacking representation, and Fig. 3b shows how the current $I$ (in µA) as a function of with time $t$ (in s). The applied voltage $V$ (in V) was increased from 0 to -2.2 V during the first 60 s of the experiment, after which $V$ was kept constant at -2.2 V. The current was relatively constant apart from a small increase of a factor of two and apparent noise (Fig. 3b). A limited change in current, within 2-3 fold, suggest that the junctions retain their functional stability under the applied conditions.[17,50]

Fig. 3c shows the corresponding changes in emission intensity with time. From this graph, oscillations are apparent at different time scales. Fourier transform was performed (FT, Note S5) to evaluate the oscillatory components ($\omega$) in the emission events with a frequency range of 1-1000 mHz (Fig. 3d, Table S1). Two dominant $\omega$ were found (highest in amplitude) at 3.3 and 20 mHz . From these results, we conclude that the single emission events possess oscillatory nature. However, our results also indicate that the oscillatory behavior of the junction is irregular where different components of the oscillatory system are coupled in a time-dependent manner. To disentangle these time-dependent components, we performed more detailed analysis. First, we determine the non-stationarity of the emission in time, meaning its statistical properties, such as mean, variance and frequency content are not constant. The Augmented Dickey-Fuller (ADF) test[51,52] and the Kwiatkowski-Phillips-Schmidt-Shin (KPSS) test[53] are complementary, and are often used together to determine the stationarity of a signal (see Note S6 for more details). ADF tests the presence of a unit root in the data and consequently the non-stationarity of the dataset, while KPSS tests whether the dataset is stationary around a deterministic trend. Both tests confirm the non-stationarity of the data, suggesting that the frequency composition changes over time.

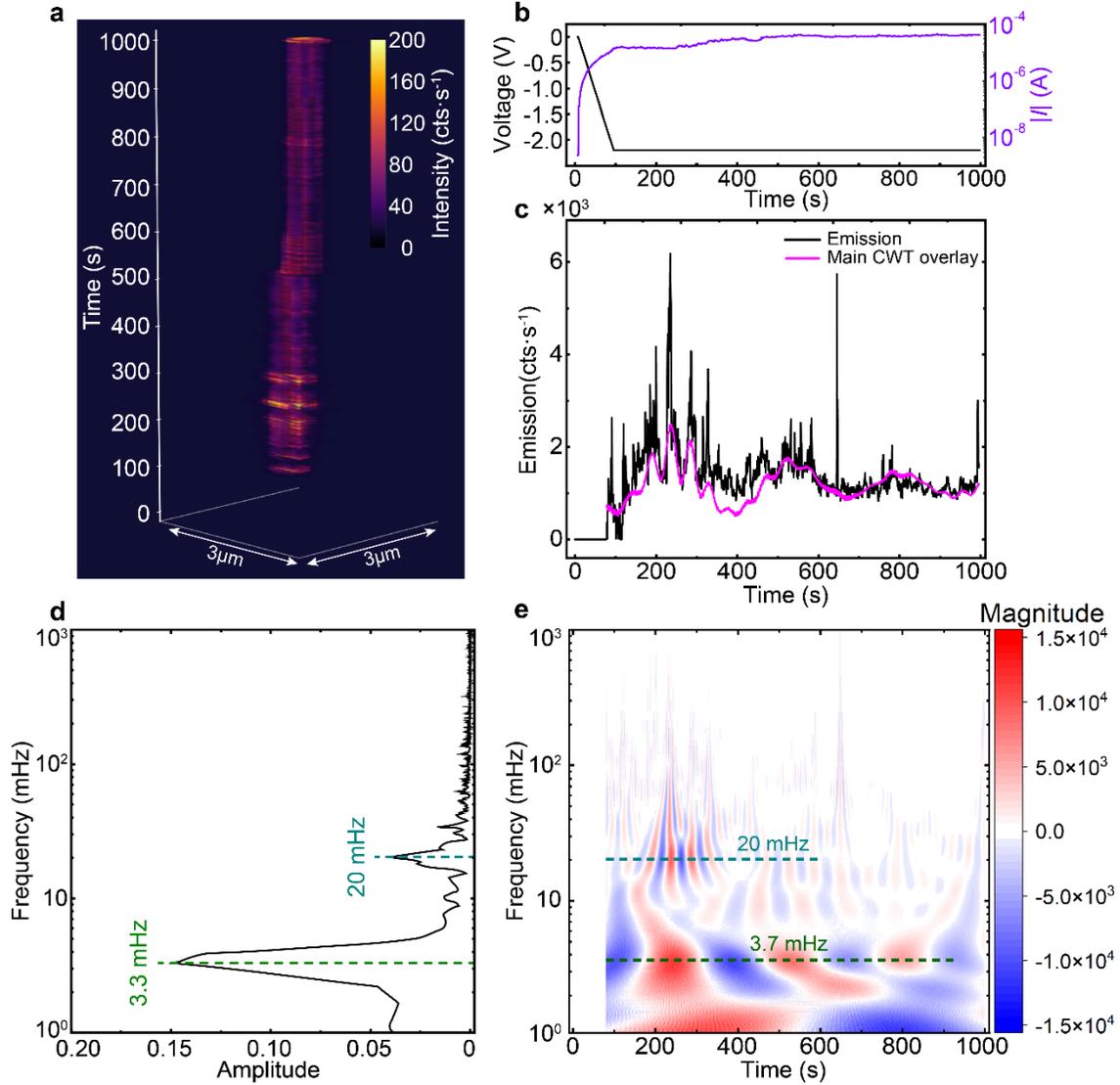

**Fig. 3 Emission and analysis of a Au-NPT//EGaIn junction at -2.2V: Oscillations of a single point source. a** Time-series EMCCD images, with pixels in *x-y* plane and time in *z* axis. Inferno color scales with emission intensity. **b** Applied voltage (in black) and measured current (in purple) plotted in time. **c** Emission intensity as function of time (black) fitted (magenta) with the two most intense oscillatory components found from d and e, confirming agreement between experiment a and analysis b, c. **d** Fourier transform (FT) analysis of emission intensity from c. **e** Continuous wavelet transform (CWT) analysis of emission intensity from a. Both d and e show two dominant oscillations at 3.7 (or 3.3 mHz) and 20 mHz.

After we identified the non-stationary character of the junctions, we characterized the temporal evolution of different $\omega$ with continuous wavelet transform (CWT, Note S7) analysis of the emission events. CWT analysis complements FT analysis by providing information on the amplitude of each $\omega$ in time (Fig. 3e), unlike FT analysis which only gives a single representation of $\omega$ over the entire time series (and therefore only works well for stationary oscillatory systems). The CWT analysis is in good agreement with FT analysis showing the two dominant $\omega$ at 3.7 and 20 mHz (Fig. 3d,e). The CWT analysis further confirms the non-stationarity of the oscillatory emission intensity with time. In other words, the frequency composition $P(\omega,t)$, *i.e.*, the $\omega$ contributing at a given time, $t$, varies dynamically. The low frequency oscillations found show a long lifetime as >800 s for 3.7 mHz compared to <600 s for 20 mHz. The two dominant $\omega$ (from FT and CWT) were used with their respective magnitude coefficients to show the agreement of the analysis with the experimental signal (magenta in Fig. 3c). Fig. 3e also reveals the high-frequency (>30 mHz) components at lower amplitude (<0.02), as characterized by shorter lifetime ($\leq$50 s) compared to the two dominant components and intermittent bursting behavior. The high-frequency components are typically short-lived episodes irregularly spaced in time, where they are active over a few oscillations before decaying. These features in the CWT analysis correspond closely to the small signals in the FT analysis. An inverse relationship between frequency, amplitude, and lifetime thus emerges: as the frequency decreases, both amplitude and lifetime increase. This trend indicates that lower-frequency components predominate, as a consequence of their higher amplitudes, dominating the oscillatory behavior of the junctions. The results from FT and CWT analysis demonstrate how single emission events possess a complex non-stationary oscillatory behavior, as characterized by long-lived low-frequency oscillations and transient higher-frequency components.

**Oscillations across the molecular junction.**

To investigate the temporal evolution of the overall emission intensity from the entire junction area, we collected videos with same parameters as in the previous experiment over 920 s, after which we stopped the experiment. Fig. 4a shows the same type of 3D image stacking representation as in Fig. 3a but extracted from Video S4. In this example, a few emission events are visible, but already by eye one emission spot (spot 1) is bright and lasts for the entire duration of the experiment while the other emission spots are visible till about $t$ = 750 s. Interestingly, the emission of spot 1 turns off when emission of spot 2 becomes very bright when the other emission spots turn off. These observations show the emission spots behave as parallel resistors following Kirchhoff's circuit rules, where the total voltage drop across the junction remains constant, but in- or de-crease of the voltage drop in one resistor (scales with intensity of an emission spot) is compensated by a corresponding in- or de-crease of voltage drops of the other resistors. In the next section we provide a more detailed analysis.

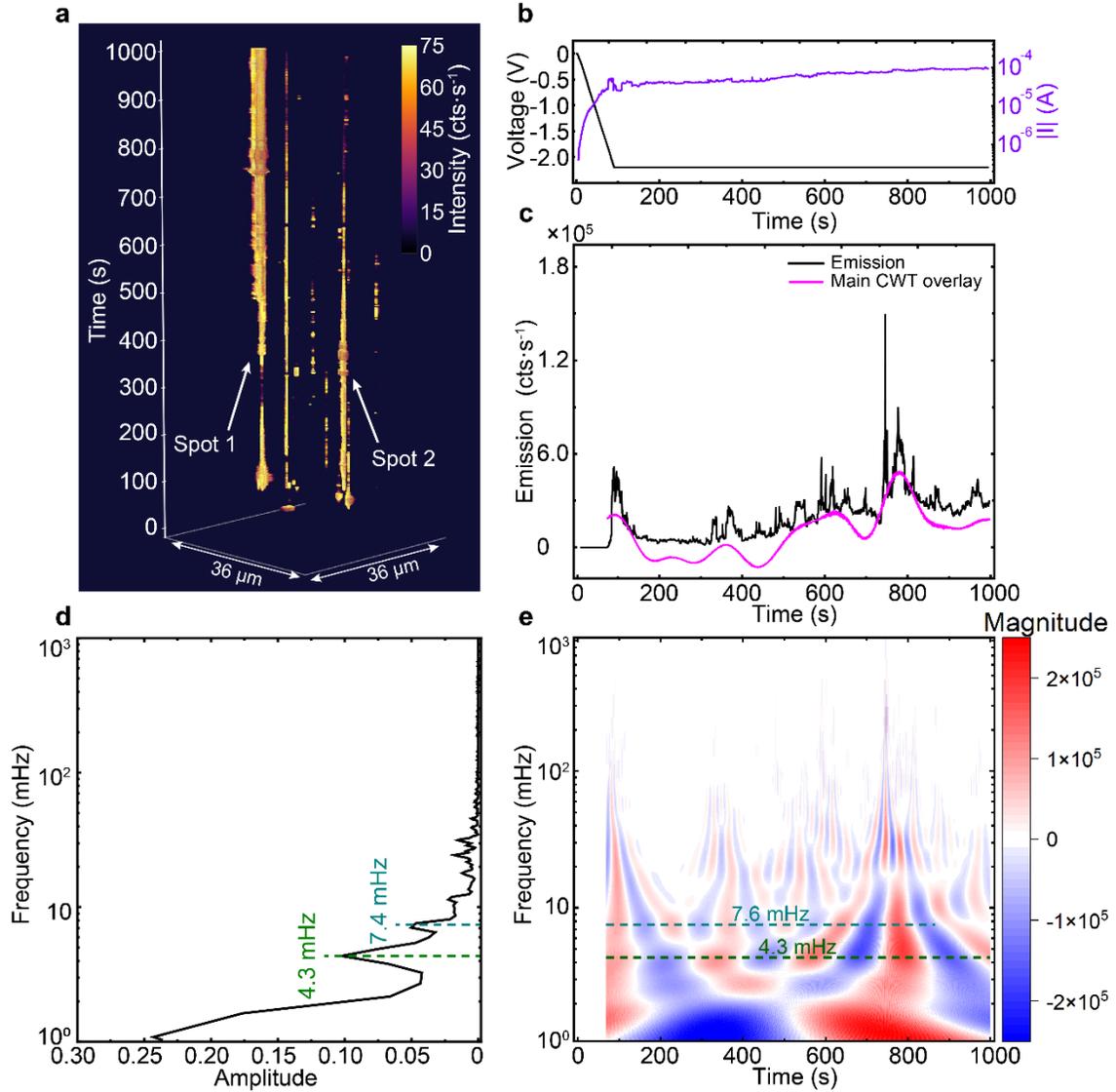

**Fig. 4 Emission and analysis of another Au-NPT//EGaIn junction at -2.2V: oscillations across the molecular junction. a** Time-series EMCCD images, with pixels in *x-y* plane and time in *z* axis. Inferno color scales with emission intensity. **b** Applied voltage (in black) and log-current (in purple) as function of time. **c** Emission intensity as function of time (black) fitted (magenta) with the two most intense oscillatory components found from d and e, confirming agreement between experiment a and analysis b, c. **d** FT analysis of emission intensity from c. **e** CWT analysis of emission intensity from a. Both d and e show two dominant oscillations at 4.3 and 7.6 (or 7.4) mHz.

Fig. 4b shows $I$(t) curve obtained when $V$ was ramped up to $V$ = -2.2 V in the first 60 s, after which $V$ was kept constant. Time $t$ = 0 s is defined at which $V$ was kept constant. Also, for this junction the current increased exponentially with $V$ but then gradually increased by a factor two. Fig. 4c shows the emission intensity as a function of time. The relevant FT (Fig. 4d) shows two highest-amplitude $\omega$ at 4.3 and 7.4 mH, from which we conclude that for this junction the overall emission is also oscillatory. ADF and the KPSS tests also confirm the non-stationarity of the oscillations of this junction (Note S3). CWT analysis shows good agreement with the FT analysis, where the two dominant $\omega$ (with highest magnitudes) at 4.3 and 7.6 mHz last for 920 and 800 s, respectively (Fig. 4e). The two dominant $\omega$ (from FT and CWT) were used with their respective magnitude coefficients to show the agreement of the analysis with the experimental signal (magenta in Fig. 4c). Fig. 4e also shows the intermittent bursting behavior at higher frequencies (>30 mHz) characterized by shorter lifetimes ($\leq$ 50 s) compared to the two dominant components.

The total junction emission, defined as the sum of all individual emission spots, also exhibits non-stationary oscillatory behavior, indicating that this phenomenon is not limited to single spots. Thus, the oscillations are dominated by the long-lived, low-frequency, and highest-amplitude components, while the high-frequency components are responsible for the bursting sharp emission spikes.

**Correlated oscillatory emission between spots.**

As explained above and visualized in Fig. 1, the emission spots are correlated in temporally oscillatory behavior. To demonstrate the correlation, and whether it is positive or negative, we carried out a more detailed analysis. A positive correlation is expected between emission events that vary in the same direction over time (both increasing or both decreasing), whereas a negative correlation is expected when their temporal trends are opposite. Fig. 5a and 5b show two successive EMCCD frames recorded from a stable junction from timestamp 0 to 0.5 s. Here, magenta and cyan circles indicate spots where the emission intensity increases and decreases respectively, in the subsequent frame. For instance, spots 1, 5 and 8 exhibit an increase in their emission intensity. Conversely, other spots display a decrease in intensity, or even turn off completely. Take spot 7 for example, it is present at timestamp 0 s but absent at 0.5 s. To elucidate the coupling between the emission events, we constructed temporal correlation matrices that quantify the relationship between intensity fluctuations of different spots by using the Spearman correlation analysis (more information in the Method and Note S5). In these matrices, red and blue elements represent positive and negative correlations, respectively (Fig. 5c). For example, in the first row for spot 1, a strong positive correlation is observed with spots 5 and 8, all of which synchronized increases in emission intensity. In contrast, the remaining spots are negatively correlated to spot 1, which display opposing (decreasing) intensity trends. A similar pattern of positive correlation is observed among emission events that exhibit simultaneous decreases in intensity. An example is given by the positive correlation between spots 3 and 4 both decreasing in emission intensity.

Fig. 5d and 5e show the emission from the same junction at the subsequent time interval from 0.5 to 1 s. During this period, spots 1, 5, and 8 continue to increase in emission intensity, while spots 2 and 9, which previously dimmed, become brighter. Conversely, spots 3, 4, and 6 exhibit continued decreases in emission, and spots 7 and 10 disappear entirely. Additionally, a new spot 11 appears. The correlation matrix for this time interval is shown in Fig. 5f, revealing a different pattern from that in Fig.5c. Once again, positive correlations are evident among spots with similar intensity trends, and negative correlations appear between those with opposing behaviors.

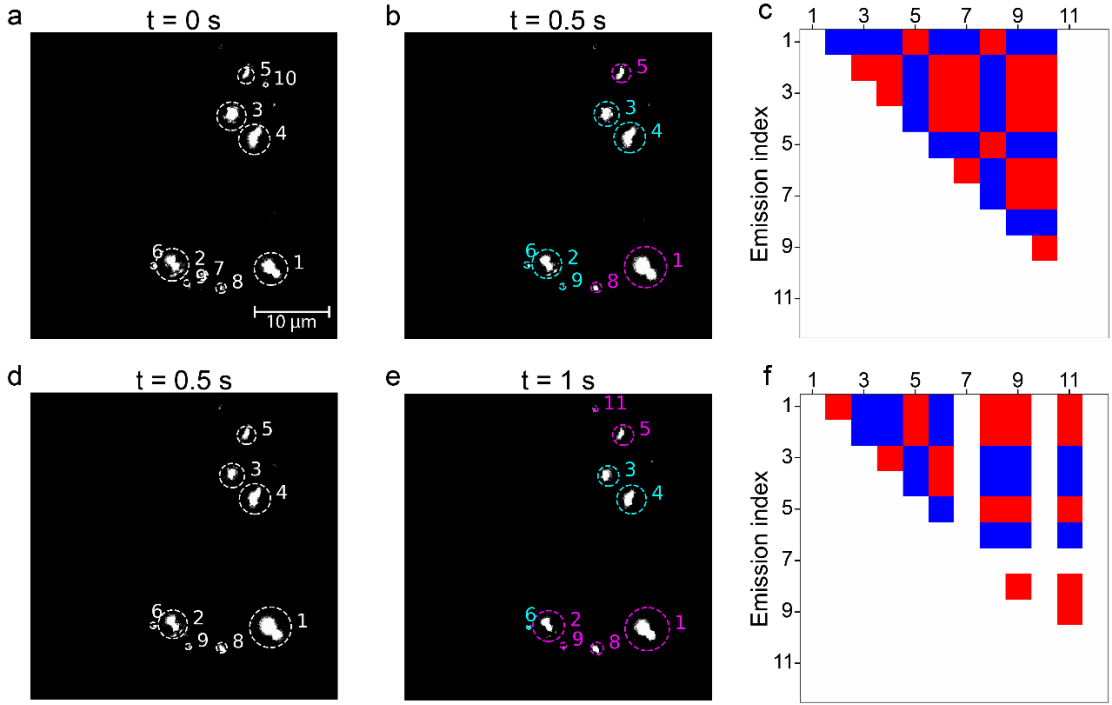

**Fig. 5 Correlation analysis of the emission spots in a stable Au-NPT//EGaIn junction. a,b** EMCCD images at timestamps $t = 0$ and 0.5 s with all the emission spots indexed and circled, where magenta indicates events with the increasing and cyan the decreasing intensity. **d,e** Similar images but at $t = 0.5$ and 1 s. **c,f** Correlation plot calculated over the two frames, for intervals 0-0.5 s (a,b) and 0.5-1 s (d,e) respectively, where red represents positive correlation and blue represents negative correlations.

Comparison of the correlation matrices at different time intervals (Fig. 5c,f) highlights the non-stationary and dynamic nature of the emission network. By comparison, a similar analysis for a junction with the appearance of a particularly bright emission spot is marked by a strong negative correlation of other locations across the junction (Fig. 6a-b). These two consecutive frames show such a bright emission event (spot 10, circled in magenta) opposes with all other spots (circled in cyan) that either dim noticeably or disappear entirely. This correlation is revealed in the matrix (Fig. 6c, blue in row and column) where spot 10 exhibits a strong negative correlation with nearly all other spots (except for spots 5 and 12). In contrast, the disappearance of a previously bright spot is accompanied by a simultaneous increase in both the number and intensity of dimmer emission events throughout the junction (Fig. 6d-e). Spot 1 undergoes a pronounced decrease in emission intensity, meanwhile all other spots (exception for spot 3) turn on or increase in emission, $t = 0.5$ s. The associated correlation matrix (Fig. 6f) reveals a similar pattern as in Fig. 6c, with negative correlation (blue elements) between the extinguishing bright emission event (spot 1) and most of the dim events across the junction.

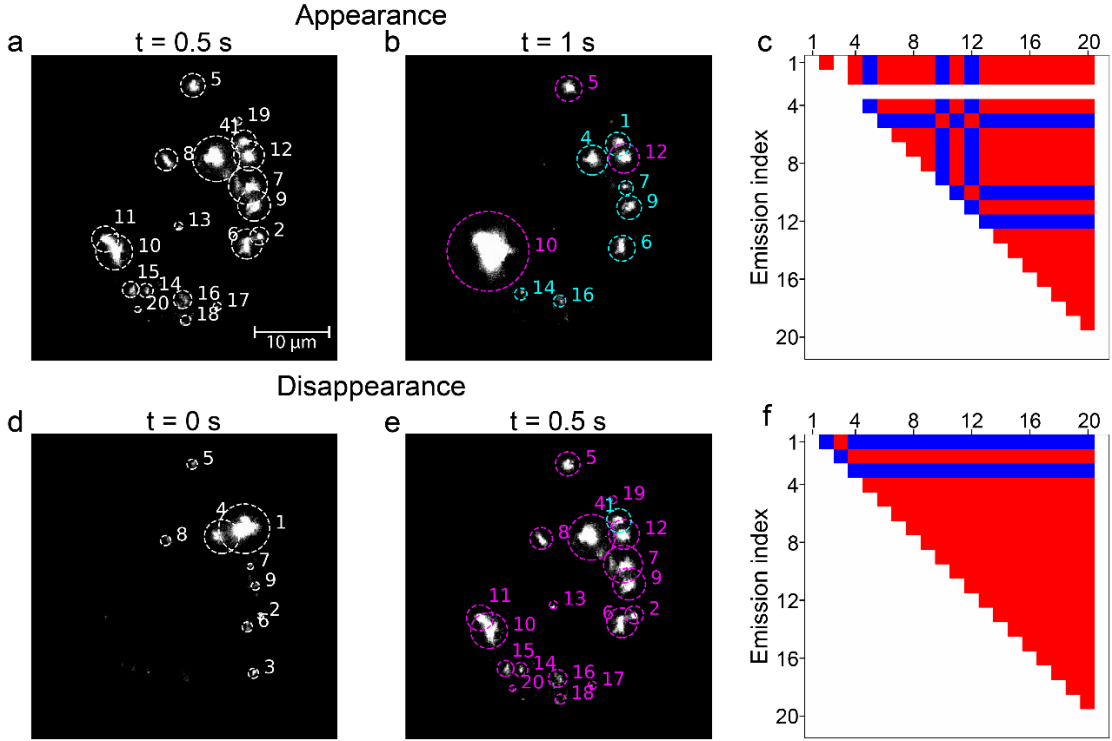

**Fig. 6 Correlation analysis of bright emission formation and disruption in a Au-NPT//EGaIn junction. a,b** EMCCD images at timestamps $t = 0.5$ and 1.0 s, showing the appearance of a bright event especially spot 10. All spots are indexed and circled, where magenta and cyan indicates events with increasing and decreasing intensity, respectively. **d,e** Similar images but at $t = 0$ and 0.5 s, showing the disappearance of the bright spot 10. Note this earlier interval is shown after the later interval in a,b to better introduce appearance before disappearance. **c,f** Correlation plot calculated over the two frames, for intervals 0.5-1 s (a,b) and 5-0.5 s (d,e) respectively. The negative correlation between the bright spots with most of the other spots is evident.

**Conclusions**

Our findings demonstrate that the dynamic emission in molecular tunneling junctions, arising from plasmons excited by inelastic tunneling, are correlated exhibiting self-oscillatory behavior. The single spots show temporal positive and negative correlations, leading to the time-dependent self-oscillations of the junction, which are characterized by the long-lasting low-frequency and transient high-frequency components. These spots behave as resistors and follow Kirchhoff's circuit rules providing a coupling mechanism where a change in current (or voltage drop) due conformational changes (as sketched in Fig. 1) from one spot has to be compensated by the other spots. Furthermore, we show that correlation map is a promising tool for probing the dynamic relationships between individual spots, offering real-time insights into molecular-scale charge transport and plasmon induced emission.

The ability to produce oscillatory emission from a static input bias, which is tuneable in frequency and general behaviour by using different molecular moieties, is relevant for other types of plasmonic junctions (such as nanoparticle-on-mirror geometry) that also show rich picocavities dynamics from applications in SERS and TERS. For example, our junctions and picocavities share similar behaviour for three reasons. First, the lifetime distribution of emission spots (Fig. S5) is similar to picocavities[25] that the majority are "fast" events with lifetime of <5 s, while a minority are "slow" events. Second, formation time of emissions (*i.e.,* time needed to trigger the events, Video S2 and S3) is ~10 s, which is similar in picocavities[25]. Third, the estimated electrical energy applied to the junctions is confirmed to be sufficient to overcome the energy barrier for adatom formation (Note S8), following a similar calculation from picocavities[25].

To conclude, the relation between emission spots with picocavities hopefully encourages further developments on relevant in-situ techniques. Besides potential in sensing applications, we believe that our findings offer opportunities for realizing molecular-scale optoelectronic oscillators with complex behaviour akin to natural systems which may be potentially interesting in applications where higher order complexity is desirable *e.g.,* nanoscale oscillatory light sources, and oscillatory neural network for neuromorphic inspired computing.

**Materials and methods**

**Preparation of ultra-flat template-stripped gold substrates (Au$^{TS}$).** The substrates were fabricated using template-stripping.[16,17,33] A 30 nm gold layer was deposited on the silicon wafers at the rate of 0.03-0.035 nm/s (TCO TOPdamper metal evaporator). Standard microscope glass coverslips (~170 μm thick) are cut in 2 × 1 cm shape and washed with soapy water, milliQ water, ethanol and isopropanol, and sonicate for 5 minutes between washings with different solvents. The glass substrates were then kept in isopropanol until use, when they were dried with nitrogen and glued onto the gold surface of Au$^{Si}$ wafers (Norland Optical Adhesive 61). The glue was cured for 30 min under a UV light source (Bluepoint2 EasyCure Honle UV Technology) and left for 2 h in ambient conditions before use. The template stripping of the ultra-flat substrates was then performed by wedging the side of the glued glass coverslips using a scalpel immediately before functionalization. The substrates for atomic force AFM characterization (see below and Fig. S1) were prepared in the same way except that standard glass slides were used, and they were cut in 0.8 × 0.8 cm square due to limitation from the microscope stage holder.

**Preparation of monolayers (Au-NPT).** 2-naphthalenethiols (NPT) was purchased from Sigma-Aldrich. All monolayers were prepared using the same procedure. Absolute ethanol was first degassed by bubbling dry nitrogen for at least 2 hours. 3 mM ethanolic solutions was prepared, and 3 ml was used for each substrate. The freshly template-stripped substrates (Au$^{TS}$) were then immersed in the solution. The samples are rinsed with 2 ml ethanol, and gently dried with nitrogen.

**Atomic force microscopy (AFM) characterization of Au-NPT.** The sample was glued onto a magnetic holder and loaded onto the piezo-scanner of the AFM instrument (Bruker Multimode 8, by Bruker Corporation, Billerica, MA, USA). We used Bruker ScanAsyst-Air silicon tips on nitride levers (k = 0.4 N/m, r = 2 nm, and $f_0$ = 70 kHz). A force of 0.7 nN was applied, the scan frequency was set to ~0.4 Hz, and the peak-force amplitude kept at 50 nm. We collected 512 samples per line for 1 × 1 μm scans. The roughness and height profile analysis was performed using Nanoscope Analysis software from Bruker. Figure S1 shows the AFM images of bare template-stripped Au and surfaces supporting the SAM.

**In-situ optical and electrical measurement of Au-NPT//EGaIn.** A conical EGaIn tip used as top electrode and a template-stripped gold film was used as bottom-electrode which supported the SAM. We applied bias to the top electrode while the bottom electrode was grounded. The Au-NPT samples were mounted on Nikon Ti2-U Eclipse inverted microscope. The EGaIn electrode was contained in a 10μl Hamilton syringe (Hamilton 10 uL, gastight needle gauge 26s) mounted on a 3D-stage on the microscope. A conical-shaped tip was created before every experiment to yield a junction with an area of 300-500 μm.[32,54] A 100× oil immersion objective (Nikon CFI Apochromat TIRF, NA=1.49) was used. A D-LEDI LED light source (385 nm, 475nm, 550 nm and 621 nm) was used to illuminate the Au-NPT//EGaIn junction to align sample in the focal plane. Illumination was then turned off to capture light emission from Au-NPT//EGaIn. In the special case to detect how light emission changed with interface dynamics, minimal illumination was kept on to visualize the interface: 621 nm coupled with a clean-up filter limiting the incident effective power on sample as 0.01 μW. Images of light emission were collected from the backside of the junction through the semitransparent gold electrode, using an Andor i-Xon Ultra 897 EMCCD camera (readout rate 17MHz at 16-bit). A Keithly 6430 Sub-femtoamp remote sourcemeter® was connected to both the working and counter electrodes (Hamilton syringe containing EGaIn). The experiments were performed in controlled atmosphere (humidity between 20 and 50%) and the bias was slowly increased until the set value for the experiment. The measurements were conducted at -2.2 V with the same experimental parameters (2 Hz frame rate, 2000 frames for a total of 1000s recording per measurement). Plasmon emission spectra were collected using an Andor Kimera 328i spectrograph coupled with and Andor iDus 420A-BR-DD CCD camera (readout rate 100kHz at 16-bit). The measurements were conducted at -1.5 to -2.3 V, and to increase stability for spectral acquisition, we used EGaIn top electrode constrained in polydimethylsiloxane (PDMS) stamping devices,[15,17,39] and using the same experimental parameters (5s integration time, 15 spectra for each applied bias).

**Automated analysis algorithm.** This was developed in-house to analyze the emission events from the measured junctions. The automated analysis was used to determine the emission intensity and the number of emission events in time. The events were indexed and further analyzed to obtain the lifetime distribution. The Fourier transform (FT, Note S2) calculations were then performed on the ensemble emission intensity with an iterative process that minimizes the zero-frequency (non-oscillatory) component in the FT by applying a baseline subtraction with a spectral range of 1-1000 mHz, dictated by our sampling rate of 2 Hz and acquisition times of maximum 1000 s. The Continuous Wavelet transform (CWT, Note S4) analysis was also performed on the baseline corrected data. Standard Morlet wavelet in a range of frequency from 1.058 to 812.5 mHz was used (compatible with the range of frequencies allowed by the acquisition parameters).

**Spearman correlation matrix.** The relationships among emission spots were determined by calculating Spearman's rank correlation coefficients for the temporal emission intensities of all pairs of events, identifying their correlations as either positive or negative[55] (Note S10). The correlation was positive when the emission intensities of both spots increased or decreased simultaneously, and negative when one increased while the other decreased.

**Acknowledgements**
This work was supported by the Dutch Research Council (NWO) under Vici program (VI.C.222.037) and XS program (OCENW.XS24.2.142 and 24.4.328). We acknowledge MESA+ Nanolab for equipment access to support the samples fabrications.

**Author contributions**
R.Z., C.A.N. and Q.L. conceived and designed the experiments. R.Z. performed the sample fabrications and measurements. C.A.N. and Q.L. supervised the project. R.Z., C.A.N. and Q.L. analyzed the data and wrote the manuscript with input from Z.W.

**Data availability**
All data needed to evaluate the conclusions are present in the paper and/or the Supplementary Information. Raw data are not publicly available at this time but may be obtained from the authors upon reasonable request.

**Conflict of interest**
The authors declare no competing interests.

**Supplementary information**
Supplementary material for this article is available from the journal website.

**References**

1  Adamantidis AR, Gutierrez Herrera C, Gent TC. Oscillating circuitries in the sleeping brain. *Nat Rev Neurosci* 2019; **20**: 746–762.
2  Crunelli V, Hughes SW. The slow (<1 Hz) rhythm of non-REM sleep: a dialogue between three cardinal oscillators. *Nat Neurosci* 2010; **13**: 9–17.
3  Novák B, Tyson JJ. Design principles of biochemical oscillators. *Nat Rev Mol Cell Biol* 2008; **9**: 981–991.
4  Csaba G, Porod W. Coupled oscillators for computing: A review and perspective. *Appl Phys Rev* 2020; **7**.
5  ter Harmsel M, Maguire OR, Runikhina SA, Wong ASY, Huck WTS, Harutyunyan SR. A catalytically active oscillator made from small organic molecules. *Nature* 2023; **621**: 87–93.
6  Baltussen MG, de Jong TJ, Duez Q, Robinson WE, Huck WTS. Chemical reservoir computation in a self-organizing reaction network. *Nature* 2024; **631**: 549–555.
7  Cybulski O, Garstecki P, Grzybowski BA. Oscillating droplet trains in microfluidic networks and their suppression in blood flow. *Nat Phys* 2019; **15**: 706–713.
8  Vantomme G, Elands LCM, Gelebart AH, Meijer EW, Pogromsky AY, Nijmeijer H *et al.* Coupled liquid crystalline oscillators in Huygens' synchrony. *Nat Mater* 2021; **20**: 1702–1706.
9  Zhang H, Zeng H, Eklund A, Guo H, Priimagi A, Ikkala O. Feedback-controlled hydrogels with homeostatic oscillations and dissipative signal transduction. *Nat Nanotechnol* 2022; **17**: 1303–1310.
10 Todri-Sanial A, Delacour C, Abernot M, Sabo F. Computing with oscillators from theoretical underpinnings to applications and demonstrators. *npj Unconventional Computing* 2024; **1**: 14.
11 Kumar A, Chaurasiya AK, González VH, Behera N, Alemán A, Khymyn R *et al.* Spin-wave-mediated mutual synchronization and phase tuning in spin Hall nano-oscillators. *Nat Phys* 2025; **21**: 245–252.
12 Manetakis K. Self-Sustained Oscillators. In: *Topics in LC Oscillators*. Springer, Cham, 2023, pp 19–38.
13 Jenkins A. Self-oscillation. *Phys Rep* 2013; **525**: 167–222.
14 Atkins P, De Paula J, Keeler J. *Atkins' physical chemistry*. 11th ed. Oxford University Press, 2025.
15 Wang M, Wang T, Ojambati OS, Duffin TJ, Kang K, Lee T *et al.* Plasmonic phenomena in molecular junctions: principles and applications. *Nat Rev Chem* 2022; **6**: 681–704.
16 Du W, Wang T, Chu H-S, Wu L, Liu R, Sun S *et al.* On-chip molecular electronic plasmon sources based on self-assembled monolayer tunnel junctions. *Nat Photonics* 2016; **10**: 274–280.
17 Wang T, Du W, Tomczak N, Wang L, Nijhuis CA. In Operando Characterization and Control over Intermittent Light Emission from Molecular Tunnel Junctions via Molecular Backbone Rigidity. *Advanced Science* 2019; **6**.

18　Lee J, Crampton KT, Tallarida N, Apkarian VA. Visualizing vibrational normal modes of a single molecule with atomically confined light. *Nature* 2019; **568**: 78–82.

19　Urbieta M, Barbry M, Zhang Y, Koval P, Sánchez-Portal D, Zabala N *et al.* Atomic-Scale Lightning Rod Effect in Plasmonic Picocavities: A Classical View to a Quantum Effect. *ACS Nano* 2018; **12**: 585–595.

20　Benz F, Schmidt MK, Dreismann A, Chikkaraddy R, Zhang Y, Demetriadou A *et al.* Single-molecule optomechanics in "picocavities". *Science* 2016; **354**: 726–729.

21　Sun S, Miscuglio M, Ma X, Ma Z, Shen C, Kayraklioglu E *et al.* Induced homomorphism: Kirchhoff's law in photonics. *Nanophotonics* 2021; **10**: 1711–1721.

22　Chikkaraddy R, de Nijs B, Benz F, Barrow SJ, Scherman OA, Rosta E *et al.* Single-molecule strong coupling at room temperature in plasmonic nanocavities. *Nature* 2016; **535**: 127–130.

23　Tan SF, Wu L, Yang JKW, Bai P, Bosman M, Nijhuis CA. Quantum Plasmon Resonances Controlled by Molecular Tunnel Junctions. *Science* 2014; **343**: 1496–1499.

24　Ding S-Y, Yi J, Li J-F, Ren B, Wu D-Y, Panneerselvam R *et al.* Nanostructure-based plasmon-enhanced Raman spectroscopy for surface analysis of materials. *Nat Rev Mater* 2016; **1**: 16021.

25　Lin Q, Hu S, Földes T, Huang J, Wright D, Griffiths J *et al.* Optical suppression of energy barriers in single molecule-metal binding. *Sci Adv* 2022; **8**: 9285.

26　Xie Z, Bâldea I, Frisbie CD. Energy Level Alignment in Molecular Tunnel Junctions by Transport and Spectroscopy: Self-Consistency for the Case of Alkyl Thiols and Dithiols on Ag, Au, and Pt Electrodes. *J Am Chem Soc* 2019; **141**: 18182–18192.

27　Qian H, Hsu S-W, Gurunatha K, Riley CT, Zhao J, Lu D *et al.* Efficient light generation from enhanced inelastic electron tunnelling. *Nat Photonics* 2018; **12**: 485–488.

28　Chen X, Roemer M, Yuan L, Du W, Thompson D, del Barco E *et al.* Molecular diodes with rectification ratios exceeding $10^5$ driven by electrostatic interactions. *Nat Nanotechnol* 2017; **12**: 797–803.

29　Akkerman HB, Blom PWM, de Leeuw DM, de Boer B. Towards molecular electronics with large-area molecular junctions. *Nature* 2006; **441**: 69–72.

30　Reus WF, Thuo MM, Shapiro ND, Nijhuis CA, Whitesides GM. The SAM, Not the Electrodes, Dominates Charge Transport in Metal-Monolayer//$Ga_2O_3$/Gallium–Indium Eutectic Junctions. *ACS Nano* 2012; **6**: 4806–4822.

31　Jiang L, Sangeeth CSS, Wan A, Vilan A, Nijhuis CA. Defect Scaling with Contact Area in EGaIn-Based Junctions: Impact on Quality, Joule Heating, and Apparent Injection Current. *The Journal of Physical Chemistry C* 2015; **119**: 960–969.

32　Chen X, Hu H, Trasobares J, Nijhuis CA. Rectification Ratio and Tunneling Decay Coefficient Depend on the Contact Geometry Revealed by in Situ Imaging of the Formation of EGaIn Junctions. *ACS Appl Mater Interfaces* 2019; **11**: 21018–21029.

33  Du W, Han Y, Hu H, Chu H-S, Annadata H V., Wang T *et al.* Directional Excitation of Surface Plasmon Polaritons via Molecular Through-Bond Tunneling across Double-Barrier Tunnel Junctions. *Nano Lett* 2019; **19**: 4634–4640.

34  Du W, Chen X, Wang T, Lin Q, Nijhuis CA. Tuning Overbias Plasmon Energy and Intensity in Molecular Plasmonic Tunneling Junctions by Atomic Polarizability. *J Am Chem Soc* 2024; **146**: 21642–21650.

35  Duffin TJ, Kalathingal V, Radulescu A, Li C, Pennycook SJ, Nijhuis CA. Cavity Plasmonics in Tunnel Junctions: Outcoupling and the Role of Surface Roughness. *Phys Rev Appl* 2020; **14**: 044021.

36  Makarenko KS, Hoang TX, Duffin TJ, Radulescu A, Kalathingal V, Lezec HJ *et al.* Efficient Surface Plasmon Polariton Excitation and Control over Outcoupling Mechanisms in Metal–Insulator–Metal Tunneling Junctions. *Advanced Science* 2020; **7**.

37  Jiang L, Sangeeth CSS, Nijhuis CA. The Origin of the Odd–Even Effect in the Tunneling Rates across EGaIn Junctions with Self-Assembled Monolayers (SAMs) of *n*-Alkanethiolates. *J Am Chem Soc* 2015; **137**: 10659–10667.

38  Chen X, Hu H, Trasobares J, Nijhuis CA. Rectification Ratio and Tunneling Decay Coefficient Depend on the Contact Geometry Revealed by in Situ Imaging of the Formation of EGaIn Junctions. *ACS Appl Mater Interfaces* 2019; **11**: 21018–21029.

39  Han Y, Nickle C, Zhang Z, Astier HPAG, Duffin TJ, Qi D *et al.* Electric-field-driven dual-functional molecular switches in tunnel junctions. *Nat Mater* 2020; **19**: 843–848.

40  Yuan L, Jiang L, Nijhuis CA. The Drive Force of Electrical Breakdown of Large-Area Molecular Tunnel Junctions. *Adv Funct Mater* 2018; **28**.

41  Chen X, Annadata H V., Kretz B, Zharnikov M, Chi X, Yu X *et al.* Interplay of Collective Electrostatic Effects and Level Alignment Dictates the Tunneling Rates across Halogenated Aromatic Monolayer Junctions. *J Phys Chem Lett* 2019; **10**: 4142–4147.

42  Li T, Bandari VK, Schmidt OG. Molecular Electronics: Creating and Bridging Molecular Junctions and Promoting Its Commercialization. *Advanced Materials* 2023; **35**.

43  Chen H, Fraser Stoddart J. From molecular to supramolecular electronics. *Nat Rev Mater* 2021; **6**: 804–828.

44  Zhang H, Zhu Y, Duan P, Shiri M, Yelishala SC, Shen S *et al.* Energy conversion and transport in molecular-scale junctions. *Appl Phys Rev* 2024; **11**.

45  Griffiths J, de Nijs B, Chikkaraddy R, Baumberg JJ. Locating Single-Atom Optical Picocavities Using Wavelength-Multiplexed Raman Scattering. *ACS Photonics* 2021; **8**: 2868–2875.

46  Baumberg JJ. Picocavities: a Primer. *Nano Lett* 2022; **22**: 5859–5865.

47  Häkkinen H. The gold–sulfur interface at the nanoscale. *Nat Chem* 2012; **4**: 443–455.

48  Baumberg JJ, Esteban R, Hu S, Muniain U, Silkin I V., Aizpurua J *et al.* Quantum Plasmonics in Sub-Atom-Thick Optical Slots. *Nano Lett* 2023; **23**: 10696–10702.

49  Carnegie C, Urbieta M, Chikkaraddy R, de Nijs B, Griffiths J, Deacon WM *et al.* Flickering nanometre-scale disorder in a crystal lattice tracked by plasmonic flare light emission. *Nat Commun* 2020; **11**: 682.

50  Zhu S, Li X, Xu W, Guo Q, Du W, Wang T. Revealing the correlation relation between conducting channels in self-assembled monolayer tunnel junctions. *J Mater Chem C Mater* 2024; **12**: 7103–7109.

51  Dickey DA, Fuller WA. Distribution of the Estimators for Autoregressive Time Series With a Unit Root. *J Am Stat Assoc* 1979; **74**: 427.

52  Dickey DA, Fuller WA. Likelihood Ratio Statistics for Autoregressive Time Series with a Unit Root. *Econometrica* 1981; **49**: 1057.

53  Kwiatkowski D, Phillips PCB, Schmidt P, Shin Y. Testing the null hypothesis of stationarity against the alternative of a unit root. *J Econom* 1992; **54**: 159–178.

54  Chiechi RC, Weiss EA, Dickey MD, Whitesides GM. Eutectic Gallium–Indium (EGaIn): A Moldable Liquid Metal for Electrical Characterization of Self-Assembled Monolayers. *Angewandte Chemie International Edition* 2008; **47**: 142–144.

55  Spearman C. The Proof and Measurement of Association between Two Things. *Am J Psychol* 1904; **15**: 72.

# Self-Oscillatory Light Emission in Plasmonic Molecular Tunnel Junctions

Riccardo Zinelli[a], Zijia Wu[a], Christian A. Nijhuis[a,*], Qianqi Lin[a,*]

[a] Hybrid Materials for Opto-Electronics Group, Department of Molecules and Materials, MESA+ Institute for Nanotechnology and Center for Brain-Inspired Nano Systems, Faculty of Science and Technology, University of Twente, P.O. Box 2017, 7500 AE Enschede, The Netherlands

*Author to whom correspondence should be addressed:
c.a.nijhuis@utwente.nl and q.lin@utwente.nl

**Note S1 Atomic force microscopy characterization of the surfaces**

The samples were glued onto a magnetic holder and loaded onto the piezo-scanner of the AFM instrument (Bruker Multimode 8, by Bruker Corporation, Billerica, MA, USA). We used Bruker ScanAsyst-Air silicon tips on nitride levers (k = 0.4 N/m, r = 2 nm, and $f_0$ = 70 kHz). A force of 0.7 nN was applied, the scan frequency was set to ~0.4 Hz, and the peak-force amplitude kept at 50 nm. We collected 512 samples per line for 1 × 1 μm scans. The roughness and height profile analysis was performed using Nanoscope Analysis software from Bruker.

We tested the surface quality of template stripped Au electrode (with a gold thickness of 30 nm; Fig. S1a), and the naphthalene-2-thiol self-assembled monolayer (Fig. S1b). The images show clean surfaces, with visible gold grains, no significant aggregated particles and only a slight increase in the surface roughness from 0.33 nm of the gold electrodes to 0.42 nm of the naphthalene-2-thiol SAM.

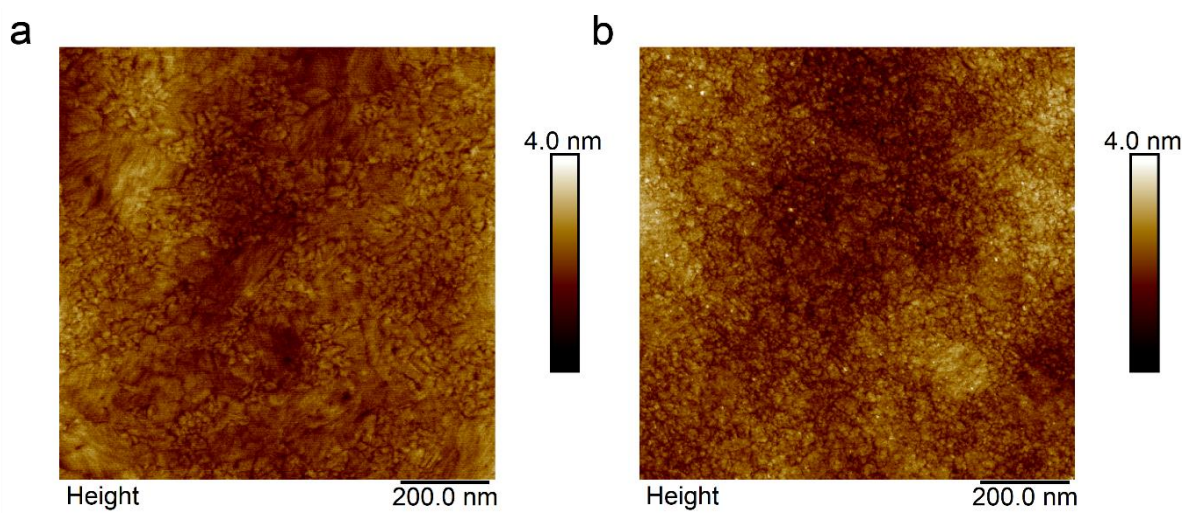

**Fig. S1 AFM images of (a) bare Au$^{TS}$ surface and (b) NPT monolayer.** The root-mean-squared (rms) surface roughness of 0.33 nm (**a**) and 0.42 nm (**b**) was determined over an area of 1 × 1 μm$^2$.

**Note S2 Photon cutoff energy for the spectra**

We measured the high energy photon cutoff energy from the plasmonic emission spectra, and we determined the high photon cutoff energy as previously reported.[1] Figure S2 shows the photon cutoff energy as a function of $V$ = applied voltage showing a blue shift in the high energy photon cutoff with increasing bias as expected from the quantum cutoff law $h\nu < eV$ ($h$ = Planck's constant, $\nu$ = photon energy, and $e$ = elementary charge). A small overbias emission can be seen for the lower biases as already observed in other techniques as well.[2–4]

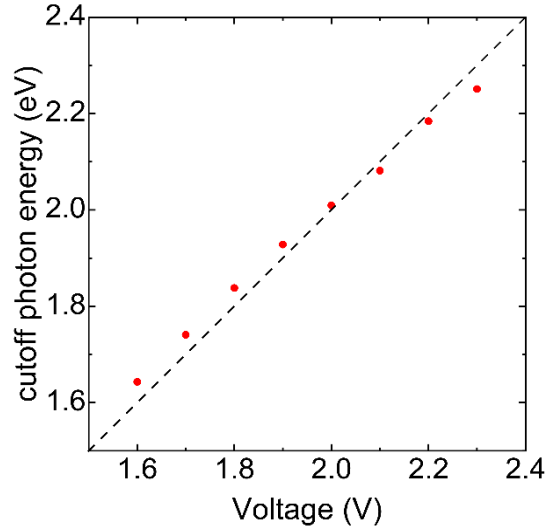

**Fig. S2 Photon cutoff energy**. Plotted *vs* applied voltage determined from the spectra in Fig. 1d.

**Note S3 Electrical characteristization of SAM-based tunnel junctions**

To perform the electrical characterization of STJs, SAMs of naphthalene-2-thiolate on Au were prepared. We recorded the $J(V)$ characteristics in the range -0.50 V to +0.50 V with steps of 50 mV. The values of $\log|J|$ for each measured bias were plotted in histograms to which we fitted a Gaussian (Fig. S3 shows 3 examples at -0.5 V, 0 V, and +0.5 V) to obtain the Gaussian log-mean, $<\log|J|>_G$, and the Gaussian log-standard deviation of $J$ which where used to plot Fig. 1c in the main text following previously reported methods.[5]

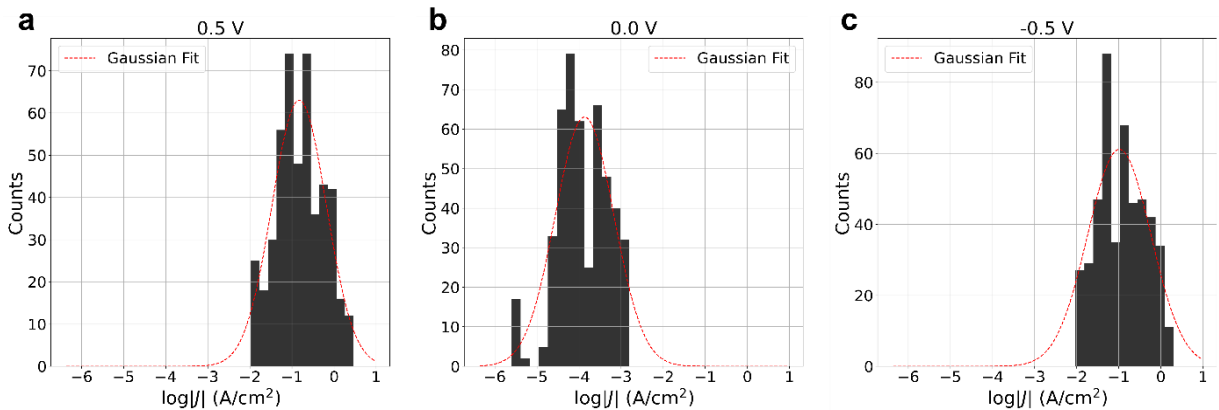

**Fig. S3 Histograms and Gaussian fits of $\log_{10}|J|$ for Au-NPT//EGaIn junctions.** $\log_{10}|J|$ distribution of 26 junctions for a total of 500 scans at (**a**) 0.5V, (**b**) 0 V, and (**c**) -0.5 V. The red dashed lines are the Gaussian fitting curves.

## Note S4 Bright emission events and shorting junctions

As described in the main text, plasmonic emission has been observed from shorting junctions. Here, a short means the formation of a metal filament leading to catastrophic failure as described in detail elsewhere.[6] When a junction (Fig.S4b) becomes unstable and breaks down (Fig.S4c), the emission behavior changes, with the emission spots turning off. The short initial position can be observed as a localized darkening of the interface (indicating alloying of EGaIn top electrode with the Au bottom electrode) and a change in the contact morphology (circled in red). The local junction failure then expands to the whole surface, usually accompanied by bright short-lived emission events, which could be associated to the formation of transient filament in the shorting regions of the junction (Fig S4d).

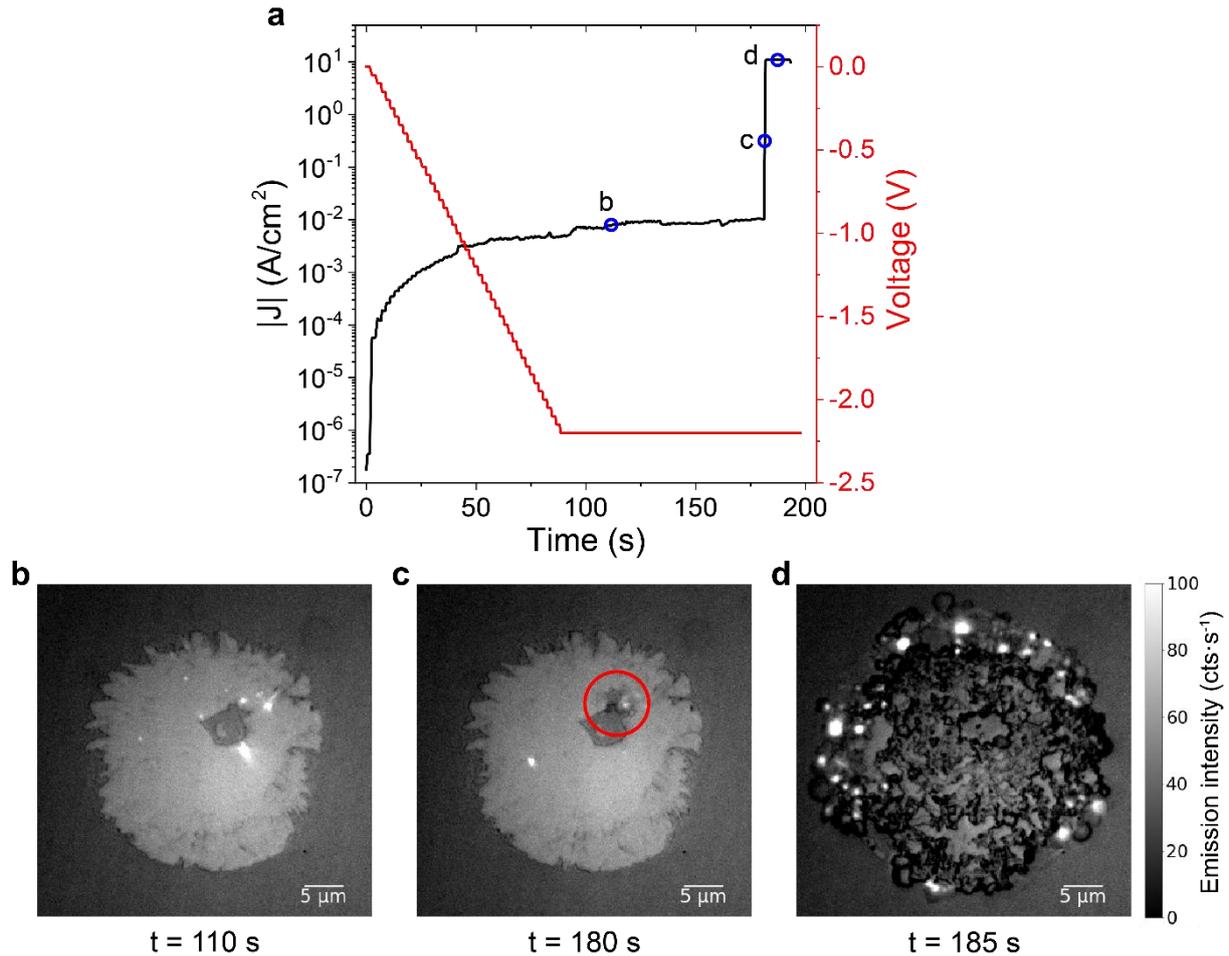

**Fig. S4 Shorting Au-NPT//EGaIn junction from Video S2**. **a** Voltage and current over time, noting timestamps for b-d. **b** Frame at $t$ =110 s, showing emission from multiple discrete spots. **c** frame at t = 180 s, showing the junction starting to fail. At the same time a spike appears in the current (a, point c). red circle shows the place where the junction start to short, visible as a darker structure in the junction **d** Frame at $t$ = 185 s, showing the complete breakdown of the junction with bright emission in the failing regions. Current reaches instrumental compliance limit which was set at 105 mA (a, point d).

## Note S5 Fourier transform (FT) of emission from Au-NPT//EGaIn junctions at different biases

Emissions were measured at applied bias of -2.2 (Table S1). The autocorrelation, after baseline correction (to remove the spurious noise), is then Fourier transformed. The main frequencies ($\omega$) from samples (*S*) and junctions (*J*) are then extracted, with their amplitudes and phases (Table 1).

**Table S1 Emission frequencies of Au-NPT//EGaIn junctions at -2.2 V.**

| Voltage (V) | Junction | Frequency (mHz) | Amplitude | Phase (°) | Voltage (V) | Junction | Frequency (mHz) | Amplitude | Phase (°) |
|---|---|---|---|---|---|---|---|---|---|
| -2.2 | S01J06 | 1.08 | 0.158734 | 360 | -2.2 | S01J09 | 1.1 | 0.04709 | 360 |
| | | 5.42 | 0.09756 | 360 | | | 3.3 | 0.14915 | 360 |
| | | 12.46 | 0.01714 | 179 | | | 7.69 | 0.01345 | 359 |
| | | 27.09 | 0.00843 | 358 | | | 10.44 | 0.01286 | 179 |
| | | 31.43 | 0.0115 | 357 | | | 14.28 | 0.01072 | 359 |
| | | 38.47 | 0.01184 | 177 | | | 18.13 | 0.02665 | 178 |
| | | 47.14 | 0.01233 | 176 | | | 20.32 | 0.04004 | 178 |
| | | 51.48 | 0.00757 | 175 | | | 30.76 | 0.00819 | 357 |
| | | 68.82 | 0.00891 | 174 | | | 34.06 | 0.01557 | 357 |
| | S01J10 | 1.08 | 0.17425 | 360 | | S01J21 | 2.71 | 0.03139 | 180 |
| | | 4.32 | 0.06963 | 360 | | | 6.5 | 0.09824 | 359 |
| | | 8.65 | 0.04744 | 359 | | | 13.55 | 0.03318 | 179 |
| | | 12.43 | 0.01381 | 179 | | | 27.09 | 0.01267 | 358 |
| | | 16.65 | 0.00612 | 178 | | | 31.43 | 0.01548 | 357 |
| | | 19.99 | 0.01955 | 178 | | | 49.85 | 0.00966 | 356 |
| | | 24.32 | 0.00956 | 178 | | | 57.44 | 0.01141 | 355 |
| | | 30.26 | 0.00969 | 357 | | | | | |
| | | 46.64 | 0.00552 | 356 | | | | | |
| | | 53.59 | 0.00727 | 175 | | | | | |
| | S03J07 | 2.27 | 0.12263 | 360 | | | | | |
| | | 5.1 | 0.09021 | 180 | | | | | |
| | | 7.37 | 0.0413 | 259 | | | | | |
| | | 10.2 | 0.03922 | 359 | | | | | |
| | | 13.03 | 0.01655 | 179 | | | | | |
| | | 19.84 | 0.01128 | 178 | | | | | |
| | | 49.87 | 0.0108 | 356 | | | | | |

**Note S6a ADF test**

The ADF statistic[7,8] is the t-stat of γ in a regression of first differences on lagged levels and lagged differences. In other words, it tests if the series returns to equilibrium over time (*i.e.*, stationary)

$$\Delta y_t = \alpha + \beta t + \gamma y_{t-1} + \sum_{i=1}^{p} \delta_i \Delta y_{t-1} + \varepsilon_t, \quad (S1)$$

$\Delta y_t = y_t - y_{t-1}$ is the first difference of the series. The data series, $y_t$, is differenced because ADF test checks if differencing makes the series stationary (*i.e.*, removes the unit root). If $y_t$ is stationary, the test should confirm that no differencing is needed. $\alpha$ is the intercept constant and represent the drift or mean level of the process. If $\alpha \neq 0$ the series may have a non-zero mean even after differencing. The inclusion of this parameter is dependent on the dataset. $\beta t$ represents, if present in the time series, a deterministic trend (for example, a steady upward growth). $\gamma y_{t-1}$ is the lagged level term, with $\gamma$ measuring how strong the series $y_{t-1}$ affects $\Delta y_t$. If $\gamma = 0$, which is the null hypothesis, the series is non-stationary, in other words, mean and variance change over time. Conversely if $\gamma < 0$, the series is stationary. Finally if $\gamma > 0$, the process in study exhibit an explosive dynamics, meaning that the deviations from equilibrium are amplified over time, meaning that the series diverge and become non-stationary and unstable. $\sum_{i=1}^{p} \delta_i \Delta y_{t-1}$, are extra lag-terms, which purpose is to absorb autocorrelation in the residual that could bias the test and invalidate the inference. $p$ is the number of lagged differences included. $\varepsilon_t$ is an error term to take white noise into account. Eq. S2 gives ADF

$$ADF = \frac{\hat{\gamma}}{SE(\hat{\gamma})}, \quad (S2)$$

where $\hat{\gamma}$ is estimated coefficient of $y_{t-1}$. $SE(\hat{\gamma})$ is the standard error of $\hat{\gamma}$. Based on the 2000 datapoints of our measurements, we calculated a max lag, $p = 45$ based on the rule of thumb of $p = \sqrt{n}$, where $n$ is the number of datapoints. The results of our analysis are summarized in Table S2 and Table S3 below.

**Note S6b KPSS test**

KPSS test [9] assumes the time series to be stationary and it tests against it. In KPSS tests, the time series data are decomposed as

$$y_t = r_t + \varepsilon_t, \quad (S3)$$

where $r_t$ is a deterministic trend or constant and $\varepsilon_t$ is a white noise component (stationary). The KPSS test value is calculated using the following formula

$$KPSS = \frac{1}{T^2 \hat{\sigma}^2} \sum_{t=1}^{T} S_t^2, \quad (S4)$$

$$S_t = \sum_{i=1}^{t} \hat{\epsilon}_i, \quad (S5)$$

where $S_t$ is the cumulative residuals $\hat{\epsilon}_i$, $T$ is the sample size, $\hat{\sigma}^2$ is the long-run variance of residuals used to account for autocorrelation, estimate of the long-run variance of $\hat{\epsilon}_i$ using the Newey-West estimator

$$\hat{\sigma}^2 = \hat{\gamma}_0 + 2 \sum_{j=1}^{q} \omega_j \hat{\gamma}_j, \quad (S6)$$

where $\hat{\gamma}_j$ is the lag-j autovariance, $\omega_j$ are bartlett weights that can calculated as

$$\omega_j = 1 - \frac{j}{q+1}, \quad (S7)$$

where $q$ is a truncation lag parameter dependent on the number of datapoints $n$ of the measurement

$$n_{lags} = int\left(12 \times \left(\frac{n}{100}\right)^{\frac{1}{4}}\right), \quad (S8)$$

Based on n = 2000 datapoints of our measurements, we calculated a $n_{lags}$ = 25.

Results for ADF and KPSS tests on dataset for data in Figs. 2 and 3.

**Table S2 Statistical summary of ADF and KPSS tests performed on the data of Fig. 3.**

| Test | Calculated value | γ (ADF) | Critical value (p= 1%) | p-value (calc) | p-value thres | Result | Conclusion |
|---|---|---|---|---|---|---|---|
| ADF | -3.4776 | -0.0335 | -3.9634 | 0.0418 | 0.01 | Accept null | Non-stationary |
| KPSS | 0.6887 | - | 0.2160 | <0.01 | 0.01 | Reject null | Non-stationary |

For both ADF and KPSS test the data of Fig. 3 is non-stationary, with a $\gamma = -0.0335$.

**Table S3 Statistical summary of ADF and KPSS tests performed on the data of Fig. 4.**

| Test | Calculated value | γ (ADF) | Critical value (p= 1%) | p-value (calc) | p-value thres | Result | Conclusion |
|---|---|---|---|---|---|---|---|
| ADF | -3.3202 | -0.0301 | -3.9634 | 0.0630 | 0.01 | Accept null | Non-stationary |
| KPSS | 0.3006 | - | 0.2160 | <0.01 | 0.01 | Reject null | Non-stationary |

For both ADF and KPSS test the data of Fig. 4 is non-stationary, with a $\gamma$ = -0.0301.

**Note S7 Continuous Wavelet transform (CWT) analysis of emission frequencies from Au-NPT//EGaIn junctions**

The same emissions used for FT were also processed with the CWT analysis[10,11]. CWT analysis allows constructions of frequencies ($\omega$) in terms of timestamps (i.e. temporal evolution of different frequencies), leveraging the inner product to analyze the similarity between the signal and a wavelet. A standard Morlet wavelet[11] was used for the main computation (complex exponential carrier multiplied by a Gaussian window). Apart from the real Morlet wavelet, other wavelets, both real and complex were explored as possible alternatives, which are listed below, however the real Morlet wavelet gave the most consistent results:

- Mexican hat (Ricker) wavelet:
  - $\psi(t) = \frac{2}{\sqrt{3\sigma}\sqrt[4]{\pi}} e^{-\frac{t^2}{2\sigma^2}} \left(1 - \frac{t^2}{\sigma^2}\right),$ (S9)
- Gaussian derivative wavelets:
  - $\psi(t) = C\, e^{-t^2}$; derivative from $n = 1$ to 8, (S10)
- Complex Gaussian derivative wavelets:
  - $\psi(t) = C\, e^{-it^2} e^{-t^2}$; derivative from $n = 1$ to 8, (S11)
- Complex Morlet wavelets
  - $\psi(t) = \frac{1}{\sqrt{\pi B}} e^{-\frac{t^2}{B}} e^{i2\pi Ct},$ (S12)
  - with $B$ (bandwidth) = 0.5 to 2.5 and $C$ (frequency center) = 0.5 to 2.5
- Shannon wavelet
  - $\psi(t) = \sqrt{B} \frac{\sin(\pi Bt)}{\pi Bt} e^{i2\pi Ct},$ (S13)

The real Morlet wavelet consists solely real components (differently from the family of the complex Morlet wavelets which possess also imaginary components) yielded the most consistent results. This outcome can be attributed to the fact that the real Morlet wavelet is particularly well suited for representing oscillatory processes, whereas other wavelet types are more appropriate for capturing sharp, transient features or possess a more complex structure that was unnecessary for the objectives of this study:

$$\psi_{real}(t) = \frac{1}{\sqrt[4]{\pi}} e^{-\frac{t^2}{2}} cos(\omega_0 t), \quad (S14)$$

**Note S8 Lifetime of light emission events at different biases**

To prove that our light emission events have similar behavior compared to previously published picocavities systems[12], we measured the lifetime distribution of the emission events at different biases. We fit a bi-exponential decay showing that there is a majority of fast short-lived events, while a minor component has longer lifetime.

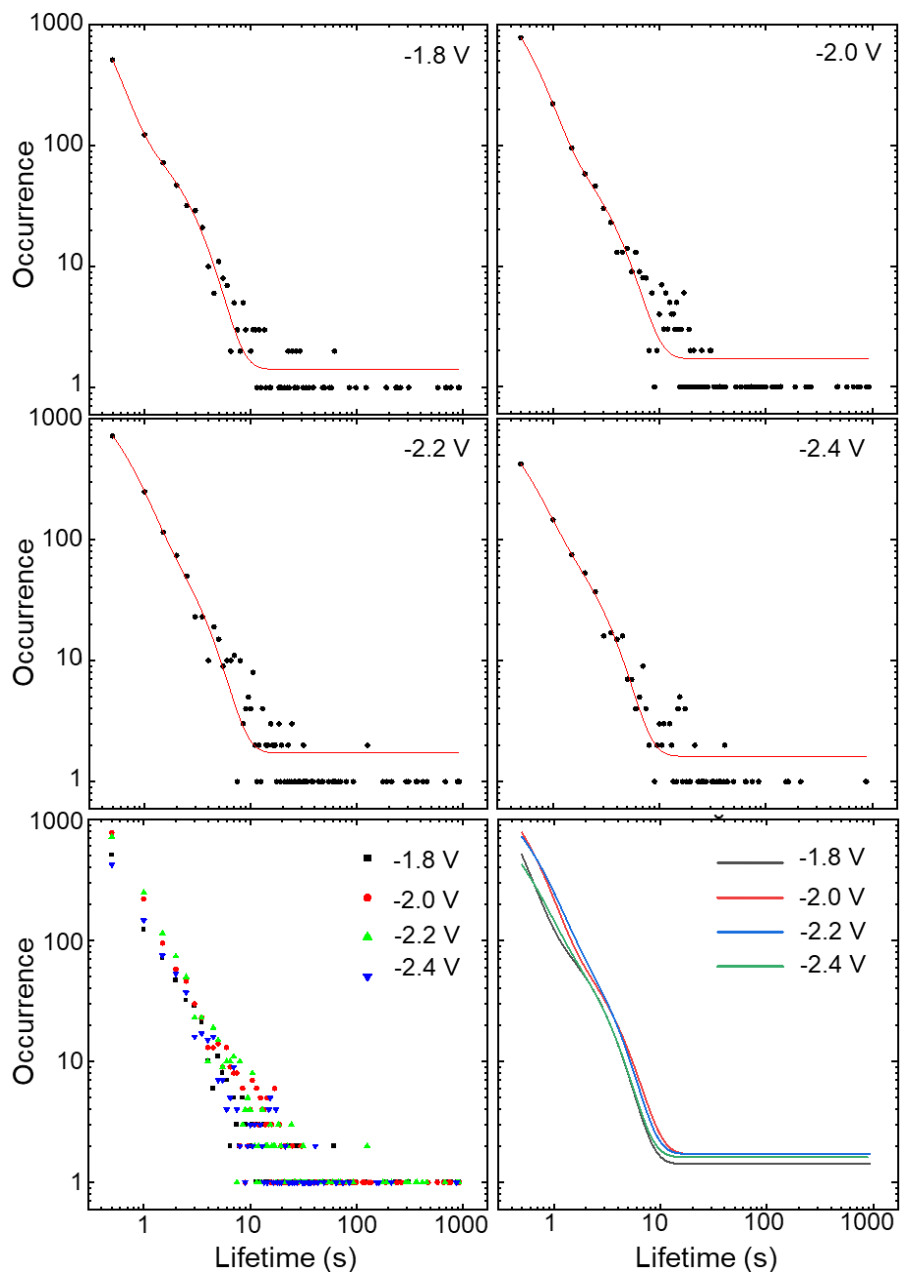

**Fig. S5 Lifetime distributions of individual emission spots at different biases.** The analysis gives occurrence distribution of individual emission lifetime, fitted with bi-exponential decay. This distribution shows lifetime is independent on the biases. Similar to "picocavities"[12], the majority of emission spots here are "fast" events (*i.e.,* short-lived, or transient) with lifetime of <5 s, while a minority are "slow" events.

**Note S9 Spearman correlation analysis**

To evaluate the correlation relationships among emission events we performed Sperman's[13] rank correlation coefficients, $r_s$, which assesses how well the relationship between two variables cane be described using a monotonic function.

$$r_s = \rho\left[R[X], R[Y]\right] = \frac{cov[R[X], R[Y]]}{\sigma_{R[X]}\sigma_{R[Y]}}, \quad (S15)$$

With $\rho$ denoting the Pearson correlation coefficient operator applied to rank variables. $cov[R[X], R[Y]]$ is the covariance of the rank variable and $\sigma_{R[X]}$ and $\sigma_{R[Y]}$ the standard deviation of the rank variables.

$$r_s = 1 - \frac{6\sum d_i^2}{n(n^2 - 1)}, \quad (S16)$$

Where $d_i$ is the difference between two ranks of each observation and n is the number of observation.

**Note S10 Estimation on electrical energy to overcome adatom formation energy barrier.**

As junction diameter is ~30 μm (Fig. 1), junction area is $\pi \times (15 \times 10^{-4})^2 = 7 \times 10^{-6}$ cm². Current at −2.2 V is ~50 μA (Fig. 3), so current density is $50 \times 10^{-6}/(7 \times 10^{-6}) = 7.1$ A/cm². Formation time[12] of emission is ~10 s (Video S2 and S3). Since most electrons tunnel elastically, IET efficiency is only $10^{-5}$.[14] For the induced plasmon, electrical energy density is thus $2.2 \times 7.1 \times 10 \times 10^{-5} = 1.6 \times 10^{-3}$ J cm$^{-2}$ = $1.0 \times 10^{16}$ eV cm$^{-2}$.

Au atomic radius is 146 pm, each bonded to at least 3 neighboring atoms.[15] For an Au atom to escape from the bulk and become adatom, it has to overcome the energy barrier to break these 3 bonds. Therefore, electrical energy transferred to this cluster is $1.0 \times 10^{16} \times \pi \times (146 \times 10^{-10})^2 \times 3 = 20$ eV.

Reported energy barrier for adatom formation from bare Au is 2 eV.[12,14] In this work, electrical energy provided at 20 eV is more than sufficient to overcome this barrier, triggering adatom formation in Au-NPT//EGaIn junctions.

**References**

1  Wang Z, Kalathingal V, Eda G, Nijhuis CA. Engineering the Outcoupling Pathways in Plasmonic Tunnel Junctions via Photonic Mode Dispersion for Low-Loss Waveguiding. *ACS Nano* 2024; **18**: 1149–1156.
2  Kalathingal V, Dawson P, Mitra J. Scanning tunnelling microscope light emission: Finite temperature current noise and over cut-off emission. *Sci Rep* 2017; **7**: 3530.
3  Du W, Wang T, Chu H-S, Wu L, Liu R, Sun S *et al.* On-chip molecular electronic plasmon sources based on self-assembled monolayer tunnel junctions. *Nat Photonics* 2016; **10**: 274–280.
4  Wang T, Du W, Tomczak N, Wang L, Nijhuis CA. In Operando Characterization and Control over Intermittent Light Emission from Molecular Tunnel Junctions via Molecular Backbone Rigidity. *Advanced Science* 2019; **6**. doi:10.1002/advs.201900390.
5  Reus WF, Nijhuis CA, Barber JR, Thuo MM, Tricard S, Whitesides GM. Statistical Tools for Analyzing Measurements of Charge Transport. *The Journal of Physical Chemistry C* 2012; **116**: 6714–6733.

6 Yuan L, Jiang L, Nijhuis CA. The Drive Force of Electrical Breakdown of Large-Area Molecular Tunnel Junctions. *Adv Funct Mater* 2018; **28**. doi:10.1002/adfm.201801710.
7 Dickey DA, Fuller WA. Distribution of the Estimators for Autoregressive Time Series With a Unit Root. *J Am Stat Assoc* 1979; **74**: 427.
8 Dickey DA, Fuller WA. Likelihood Ratio Statistics for Autoregressive Time Series with a Unit Root. *Econometrica* 1981; **49**: 1057.
9 Kwiatkowski D, Phillips PCB, Schmidt P, Shin Y. Testing the null hypothesis of stationarity against the alternative of a unit root. *J Econom* 1992; **54**: 159–178.
10 Arts LPA, van den Broek EgonL. The fast continuous wavelet transformation (fCWT) for real-time, high-quality, noise-resistant time–frequency analysis. *Nat Comput Sci* 2022; **2**: 47–58.
11 Goupillaud P, Grossmann A, Morlet J. Cycle-octave and related transforms in seismic signal analysis. *Geoexploration* 1984; **23**: 85–102.
12 Lin Q, Hu S, Földes T, Huang J, Wright D, Griffiths J *et al.* Optical suppression of energy barriers in single molecule-metal binding. *Sci Adv* 2022; **8**: 9285.
13 Spearman C. The Proof and Measurement of Association between Two Things. *Am J Psychol* 1904; **15**: 72.
14 Qian H, Hsu S-W, Gurunatha K, Riley CT, Zhao J, Lu D *et al.* Efficient light generation from enhanced inelastic electron tunnelling. *Nat Photonics* 2018; **12**: 485–488.
15 Thompson D, Liao J, Nolan M, Quinn AJ, Nijhuis CA, ODwyer C *et al.* Formation Mechanism of Metal-Molecule-Metal Junctions: Molecule-Assisted Migration on Metal Defects. *Journal of Physical Chemistry C* 2015; **119**: 19438–19451.